\documentclass[twocolumn,showpacs,preprintnumbers,amsmath,amssymb,superscriptaddress]{revtex4}

\usepackage{graphicx}
\usepackage{dcolumn}
\usepackage{bm}
\usepackage{bbm}

\usepackage{color}
\usepackage[overlay,absolute]{textpos}
\definecolor{mygray}{gray}{0.4}
\definecolor{mylink}{rgb}{0.2, 0.2, 0.5}
\usepackage[colorlinks=true,urlcolor=mylink,dvipdfm,bookmarksnumbered=true,%
pdfborder={0 0 0},pdfpagemode=UseNone,pdfstartview=FitH]{hyperref}

\begin{document}

\begin{textblock*}{\textwidth}[0,0](19mm,11.5mm)
\footnotesize\noindent 
\begin{minipage}{\textwidth}
\center
\textcolor{mygray}{Journal link:}
\href{http://dx.doi.org/10.1103/PhysRevA.81.012307}{http://dx.doi.org/10.1103/PhysRevA.81.012307}
\end{minipage}
\end{textblock*}

\begin{textblock*}{0.6\textwidth}[0,0](19mm,261mm)
\footnotesize\noindent
\begin{minipage}{\textwidth}
\textcolor{mygray}{Journal ref:} \href{http://dx.doi.org/10.1103/PhysRevA.81.012307}{A. E. B. Nielsen, Phys.\ Rev.\ A \textbf{81}, 012307 (2010)}.\\
\textcolor{mygray}{Copyright (2010) by the American Physical Society.}
\end{minipage}
\end{textblock*}

\title{Fighting decoherence in a continuous two-qubit odd or even parity measurement\\ with a closed-loop setup}

\author{Anne E. B. Nielsen}
\affiliation{Lundbeck Foundation Theoretical Center for Quantum
System Research, Department of Physics and Astronomy,
Aarhus University, DK-8000 \AA rhus C, Denmark}
\affiliation{Edward L. Ginzton Laboratory, Stanford University, Stanford CA 94305, USA}

\begin{abstract}
A parity measurement on two qubits, each consisting of a single atom in a cavity, can be realized by measuring the phase shift of a probe beam, which interacts sequentially with the two qubits, but imperfections lead to decoherence within the subspaces of a given parity. We demonstrate that a different setup, where the probe light interacts repeatedly with the qubits, can reduce the rate of decoherence within the odd or the even parity subspace significantly. We consider both the case of a resonant and the case of a nonresonant light-atom interaction and find that the performance is comparable if the parameters are chosen appropriately.
\end{abstract}

\pacs{03.67.Pp, 42.50.Pq, 03.65.Yz}
\keywords{Suggested keywords}

\maketitle

\section{Introduction}\label{I}

It has been realized that measurements provide a powerful toolbox to manipulate the state of quantum systems in ways that might be difficult to achieve with standard interactions. An example is the possibility of quantum computing with only single photon sources, linear optics, and photo detectors \cite{knill}. A measurement may consist in observing a part of a larger system such as in several recent experiments preparing various quantum states of light by conditional detection of a small part of the light field \cite{ourjoumtsev,neergaard,wakui}. Alternatively, one may use a probe to gain information about the state of a system as in atomic spin squeezing by use of Faraday rotation \cite{kuzmich}. In the latter case, the measurement should ideally be constructed in such a way that the interaction between the probe and the system does not affect the time evolution of the measured observable.

A particularly interesting case arises, when the measured observable has degenerate eigenvalues, since one may then obtain partial information about the state of the system, while leaving a specific subspace unaffected. This opens the way to use measurements to prepare macroscopic quantum superposition states \cite{massar,nm8} and to detect errors in quantum computations without destroying the quantum information stored in the qubits \cite{gottesman}. An example is the three-qubit bit-flip code, where each qubit is encoded in the two states $|000\rangle$ and $|111\rangle$ of three physical qubits, and a single bit-flip error can be detected by measuring the parity of qubits 1 and 2 and the parity of qubits 2 and 3. Other applications of parity measurements include entanglement generation \cite{williams}, quantum computing \cite{ionicioiu}, and state transfer \cite{yuan}.

Recently, it has been proposed to perform a continuous parity measurement on two qubits in a cavity quantum electrodynamics network by allowing a coherent state probe to interact sequentially with the two qubits and then measure the phase shift imposed on the probe light with a homodyne detector \cite{kerckhoff}. More precisely, each qubit is encoded in two ground state levels $|0\rangle$ and $|1\rangle$ of a single atom in a cavity, and the probe light couples the state $|1\rangle$ resonantly to an excited state $|e\rangle$. In the strong coupling limit, the vacuum Rabi splitting prevents the light from entering into the cavity if the atom is in the state $|1\rangle$ \cite{hood}, and each interaction thus leaves the probe light unaffected except for a conditional phase shift of $\pi$ \cite{duan}. The measurement takes a finite amount of time due to uncertainties arising from shot noise, and it is suitable for continuous quantum error correction, where feedback is used to keep the state of the qubits within the code space as far as possible \cite{ahn,sarovar}. Quantum error correction is often thought of as strong measurements followed by fast corrections, but continuous error correction contains some interesting aspects due to the intrinsic robustness of feedback control. In \cite{nurdin} a setup has been proposed in which the above interaction is used, but rather than detecting the probe field, the probe field is used in a feedback loop, which automatically corrects errors.

\begin{figure}
\includegraphics*[width=\columnwidth]{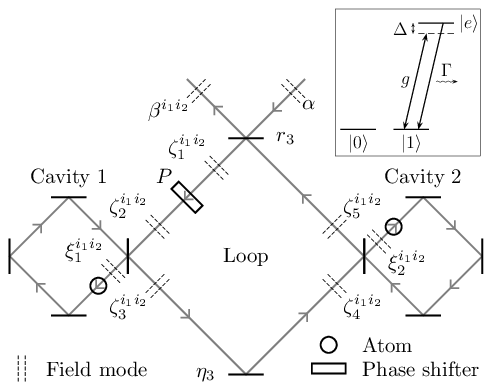}
\caption{Proposed setup to perform a parity measurement on two qubits, each encoded in two ground state levels of a single atom in a cavity. The input field $\alpha$ is in a continuous coherent state, and the output field $\beta$ is observed with a homodyne detector. The Greek letters label the field at different positions as explained in the text, $r_3^2$ is the reflectivity of the input beam splitter, $P=e^{i\psi}$ is the phase factor imposed on the light field by the phase shifter, and $1-\eta_3$ is the fraction of photons that are lost, when light travels from cavity 1 to cavity 2. The inset shows the level structure of the atoms, the coupling to the cavity field, which may be resonant ($\Delta=0$) or nonresonant ($\Delta\neq0$), and the possibility of spontaneous emission.\label{setup}}
\end{figure}

Independent of whether the probe field is detected or used for coherent feedback, it is important to minimize decoherence within the parity subspaces due to the interaction with the probe. The above scheme is sensitive to light field losses between and within the two cavities as well as to spontaneous emission from the excited state of each atom, arising because the field is not completely expelled when the coupling strength is finite. In the present paper we show that a slight change of the setup can decrease the rate of decoherence in the odd or the even parity subspace by about one order of magnitude. The basic idea is to allow the light field to interact several times with the two qubits by reflecting some of the light emerging from the second cavity back onto the input mirror of the first cavity as shown in Fig.~\ref{setup}. For the odd subspace (and $P=1$), the light field experiences a phase shift of $\pi$ for each round trip in the loop, and the resulting destructive interference between light fields having traveled a different number of round trips in the loop almost extinguishes the field. The difference between the field amplitudes obtained for the two possible states within the odd subspace is thus small, and this improves the robustness of the measurement with respect to loss. The destructive interference also decreases the field amplitudes inside the qubit cavities, which reduces the rate of spontaneous emission events and further decreases the decoherence rate. On the other hand, the sensitivity of the measurement is not decreased, since the even parity subspace leads to constructive interference in the loop, i.e., the phase of the output field differs by $\pi$ for the two subspaces as before. We note that the roles of the odd and the even parity subspaces could be exchanged by adding a phase shift of $\pi$ within the loop, but in the following we focus on protecting the odd subspace. The reason for this is that the output field is not exactly the same for the two even parity states, because spontaneous emission is possible for $|11\rangle$ but not for $|00\rangle$, and as a result the homodyne detection of the output field does not leave the even subspace completely unaffected. While this makes the analysis more complicated, it will not have any significant effects, since the corrections are of second order. A parity measurement can also be realized with a light field, which is far detuned from the atomic transition, and this possibility will also be investigated.

In Sec.~\ref{II} we provide the theoretical framework required to analyze the proposed setup, and we obtain expressions for the field amplitudes, which substantiate the above statements. In Sec.~\ref{III} we derive an equation for the time evolution of the purity of a state restricted to either the odd or the even parity subspace and identify the parameters, which characterize the performance of the proposal. These parameters are evaluated for a nonresonant light-atom interaction in Sec.~\ref{IV} and for a resonant light-atom interaction in Sec.~\ref{V}. We also compute the optimal value of the reflectivity $r_3^2$ of the input beam splitter and compare the performance of the resonant and the nonresonant interaction. Finally, Sec.~\ref{VI} concludes the paper.

\section{System state and parity measurements}\label{II}

\subsection{Light-atom interaction}\label{IIA}

To analyze the proposed setup, we first consider the interaction between a single light field mode and one atom as depicted in the inset of Fig.~\ref{setup}. The light couples the state $|1\rangle$ to the excited state $|e\rangle$ with coupling strength $g$, and the excited state decays by spontaneous emission to the state $|1\rangle$ at a rate $\Gamma$, whereas decay to the state $|0\rangle$ is assumed to be forbidden. In a frame rotating with the angular frequency $\omega$ of the light field, the master equation for the interaction takes the form
\begin{equation}\label{phatintfull}
\frac{d\rho_1}{dt}=-\frac{i}{\hbar}[H,\rho_1]+
\frac{\Gamma}{2}(2\sigma\rho_1\sigma^\dag-\sigma^\dag\sigma\rho_1
-\rho_1\sigma^\dag\sigma),
\end{equation}
where
\begin{equation}
H=\hbar g(\hat{a}\sigma^\dag+\hat{a}^\dag\sigma)+\hbar\Delta\sigma^\dag\sigma
\end{equation}
is the Hamiltonian for the interaction, $\rho_1$ is the density operator of the considered atom and the light field mode, $\sigma=|1\rangle\langle e|$ is the atomic lowering operator, $\hat{a}$ is the field annihilation operator, and $\Delta=\omega_{\textrm{at}}-\omega$ is the detuning between the atomic resonance angular frequency and the angular frequency of the light field.

We are solely interested in the parameter regime where the probability to populate the excited level is small since observation of a photon emitted by spontaneous emission from one of the two atoms allows us to distinguish the two states within the odd subspace and to distinguish the two states within the even subspace. We thus assume that either the decay rate or the absolute value of the light-atom detuning is large compared to the effective driving of the atomic transition, which amounts to the condition
\begin{equation}\label{condition}
g^2|\langle\hat{a}\rangle|^2\ll\Gamma^2/4+\Delta^2.
\end{equation}
When \eqref{condition} is satisfied, we can eliminate the excited state adiabatically. Following the procedure outlined in \cite{bouten1,bouten2}, we formally assume $g\hat{a}\propto k$, $\Gamma\propto k^2$, and $\Delta\propto k^2$ and take the limit $k\rightarrow\infty$, which transforms \eqref{phatintfull} into
\begin{multline}\label{phatint}
\frac{d\rho_1}{dt}=\frac{i\Delta g^2}{(\Gamma/2)^2+\Delta^2}
\left[\hat{a}^\dag\hat{a}|1\rangle\langle1|,\rho_1\right]
+\frac{\Gamma}{2}\frac{g^2}{(\Gamma/2)^2+\Delta^2}\\
\times(2\hat{a}|1\rangle\langle1|\rho_1|1\rangle\langle1|\hat{a}^\dag
-\hat{a}^\dag\hat{a}|1\rangle\langle1|\rho_1-\rho_1|1\rangle\langle1|\hat{a}^\dag\hat{a}),
\end{multline}
where the Hilbert space of the atom is now restricted to $\textrm{span}\{|0\rangle,|1\rangle\}$. The light-atom interaction is thus equivalent to the combined effect of a phase shifter and a light field loss if the atom is in the state $|1\rangle$, whereas no interactions take place if the atom is in the state $|0\rangle$. The loss is, in turn, equivalent to a beam splitter coupling the light mode to a vacuum mode followed by a partial trace over the latter mode. Note that there is no coupling between the states $|0\rangle$ and $|1\rangle$ in \eqref{phatint} as is desirable in a measurement scheme.

\subsection{General form of the system density operator}\label{IIB}

Having characterized the light-atom interaction, we next turn to the full setup in Fig.~\ref{setup}. In the Markov limit, where all time delays go to zero, one could derive a quantum filtering equation for the time evolution of the state of the system, for instance by using the time evolution obtained for the measurement free case in \cite{yanagisawa} and then add homodyne detection \cite{bouten3}. Here, however, we assume the approximate light-atom interaction in Eq.~\ref{phatint} and use an alternative approach, where we divide the continuous beams of light into segments of infinitesimal length and consider the transformations that occur when the segments hit different components of the setup. Each of these segments constitutes a single mode of the light field as explained in \cite{blow}.

The crucial point to note is that beam splitters, mirrors, and phase shifters all transform the light field operators linearly, i.e., for a given state of the two atoms, $|00\rangle$, $|10\rangle$, $|01\rangle$, or $|11\rangle$, all the components of the setup transform coherent states into coherent states, and furthermore homodyne detection of the output field and partial trace operations applied to the field leaking out to the surroundings do not affect the coherent amplitudes of the remaining field modes \cite{giedke}. For a continuous coherent state input field, the density operator of the state of the two atoms and the light modes within the loop and the cavities at time $t$ may hence be written as
\begin{multline}\label{state}
\rho(t)=\sum_{i_1,j_1,i_2,j_2=0}^1c_{i_1j_1i_2j_2}(t)
\bigotimes_{k=1}^{N_3}|\zeta_k^{i_1i_2}\sqrt{T_k}\rangle\langle\zeta_k^{j_1j_2}\sqrt{T_k}|\\
\bigotimes_{q=1}^2\bigotimes_{k_q=1}^{N_q}|\xi_{qk_q}^{i_1i_2}
\sqrt{\tau_{qk_q}}\rangle\langle\xi_{qk_q}^{j_1j_2}\sqrt{\tau_{qk_q}}|
\otimes|i_1i_2\rangle\langle j_1j_2|,
\end{multline}
where $i_q$ and $j_q$ specify the state of the atom in cavity $q$, $c_{i_1j_1i_2j_2}(t)$ are time dependent scalars, $N_3$ is the number of light modes in the loop, and $N_q$ is the number of light modes in cavity $q$. The $k$th mode in the loop is assumed to have temporal width $T_k$ and conditional amplitude $\zeta_k^{i_1i_2}$, normalized such that $|\zeta_k^{i_1i_2}|^2T_k$ is the expectation value of the number of photons in the $k$th mode conditioned on the atoms being in the state $|i_1i_2\rangle$. Likewise, the $k_q$th mode in cavity $q$ is assumed to have temporal width $\tau_{qk_q}$ and conditional amplitude $\xi_{qk_q}^{i_1i_2}$. The time dependence of the coefficients $c_{i_1j_1i_2j_2}(t)$, which is discussed in more detail in Sec.~\ref{III}, arises solely from the homodyne detection of the output field and the partial traces over the field lost in each time step.

\subsection{Steady state field amplitudes}\label{IIC}

In general, the conditional field amplitudes are time dependent, but if the input field is turned on at time $t=0$ and is constant for $t>0$, the amplitudes approach constant values at a rate which, in the limit $r_3\rightarrow0$, is the smallest of the decay rates of the light fields in cavity 1 and 2. In steady state, the conditional field amplitudes of adjacent modes are the same, except when crossing an optical component or one of the atoms, and it is thus sufficient to consider a small number of conditional field amplitudes. To compute the steady state values of the conditional field amplitudes of the modes shown explicitly in Fig.~\ref{setup}, we first concentrate on the transformations occurring during one round trip in cavity $q$.

It follows directly from the interpretation of Eq.~\ref{phatint} that the light-atom interaction changes the phase of the light field by $\Delta_qg_q^2/((\Gamma_q/2)^2+\Delta_q^2)$ per unit interaction time and decreases the field amplitude at the rate $(\Gamma_q/2)g_q^2/((\Gamma_q/2)^2+\Delta_q^2))$ if the atom in cavity $q$ is in the state $|1\rangle$ (the added subscripts refer to the values of the parameters in cavity $q$). We note that $g_q^2$ is inversely proportional to the mode volume of the considered mode and thus inversely proportional to the temporal width of the mode, which, in turn, is equal to the interaction time. In a cavity, $g_q$ is normally defined with respect to a mode of temporal width $\tau_q$, where $\tau_q$ is the round trip time of light in cavity $q$, and imposing this convention, the product of the square of the light-atom coupling strength and the interaction time is $g_q^2\tau_q$ independent of the actual temporal width of the mode. The three mirrors in cavity $q$ are assumed to be perfectly reflecting and to give rise to a phase factor of $i$ each. Detuning between the light field and the cavity resonance angular frequency $\omega_{c,q}$ is taken into account by multiplying the field amplitude by a factor $\exp(-i\delta_q\tau_q)$, where $\delta_q=\omega_{c,q}-\omega$. It thus follows that a single round trip in cavity $q$ from just after the input beam splitter to just before the input beam splitter transforms the conditional field amplitude from $\xi_q^{i_1i_2}$ into $-if_q^{i_q}\xi_q^{i_1i_2}$, where
\begin{equation}\label{f}
f_q^{i_q}=\exp\left(-\left(\frac{\Gamma_q}{2}-i\Delta_q\right)
\frac{g_q^2\delta_{1i_q}}{\Gamma_q^2/4+\Delta_q^2}\tau_q
-i\delta_q\tau_q\right)
\end{equation}
and $\delta_{ij}$ is the Kronecker delta.

Inspection of Fig.~\ref{setup} and the steady state conditions $\xi_1^{i_1i_2}=t_1\zeta_2^{i_1i_2}+r_1f_1^{i_1}\xi_1^{i_1i_2}$ and $\xi_2^{i_1i_2}=t_2\zeta_4^{i_1i_2}+r_2f_2^{i_2}\xi_2^{i_1i_2}$, where $r_q^2=1-t_q^2$ is the reflectivity of the input beam splitter of cavity $q$, then give $\zeta_1^{i_1i_2}=t_3\alpha+ir_3\zeta_5^{i_1i_2}$, $\zeta_2^{i_1i_2}=P\zeta_1^{i_1i_2}$, $\zeta_3^{i_1i_2}=-iF_1^{i_1}\zeta_2^{i_1i_2}$, $\zeta_4^{i_1i_2}=i\sqrt{\eta_3}\zeta_3^{i_1i_2}$, $\zeta_5^{i_1i_2}=-iF_2^{i_2}\zeta_4^{i_1i_2}$, $\beta^{i_1i_2}=t_3\zeta_5^{i_1i_2}+ir_3\alpha$,
$\xi_1^{i_1i_2}=t_1\zeta_2^{i_1i_2}/(1-r_1f_1^{i_1})$, and $\xi_2^{i_1i_2}=t_2\zeta_4^{i_1i_2}/(1-r_2f_2^{i_2})$, where $-iF_q^{i_q}=-i(f_q^{i_q}-r_q)/(1-r_qf_q^{i_q})$ describes how cavity $q$ transforms the light field in the loop in steady state. From these relations we derive
\begin{eqnarray}
\zeta_2^{i_1i_2}&=&\frac{Pt_3}{1-r_3PF_1^{i_1}F_2^{i_2}\sqrt{\eta_3}}\alpha,\label{zeta2}\\
\beta^{i_1i_2}&=&i\frac{r_3-PF_1^{i_1}F_2^{i_2}\sqrt{\eta_3}}
{1-r_3PF_1^{i_1}F_2^{i_2}\sqrt{\eta_3}}\alpha.\label{beta}
\end{eqnarray}
If the atom and cavity parameters are the same for cavity 1 and 2, it follows that $\beta^{i_1i_2}=\beta^{i_2i_1}$, i.e., the output field is exactly the same for the two two-qubit states with odd parity. This is a very important property of the system, since it ensures that it is impossible to destroy a superposition of odd parity states by doing any kind of measurements on the output field, and, in particular, inefficient detection does not give rise to decoherence within the odd subspace. In appendix~\ref{A}, we show that this is valid even during the transient and for a time dependent input field.

Since $f_q^{i_q}$ is typically very close to unity, we should choose $r_q^2$, $q=1,2$, to be close to unity as well in order to allow $F_q^{i_q}$ to be significantly different from unity. We can thus expand $f_q^{i_q}$ and $r_q$ to first order to obtain
\begin{equation}\label{xi}
\xi_1^{i_1i_2}=\frac{2}{\sqrt{\kappa_1\tau_1}}\frac{(1+D_1^2)\zeta_2^{i_1i_2}}
{1+D_1^2+2C_1\delta_{1i_1}-2iC_1D_1(\delta_{1i_1}-1/2)}
\end{equation}
and
\begin{equation}\label{fqiq}
F_q^{i_q}=\frac{1+D_q^2-2C_q\delta_{1i_q}+2iD_qC_q(\delta_{1i_q}-1/2)}
{1+D_q^2+2C_q\delta_{1i_q}-2iD_qC_q(\delta_{1i_q}-1/2)},
\end{equation}
where $C_q=2g_q^2/(\kappa_q\Gamma_q)$ is the cooperativity parameter, $\kappa_q=t_q^2/\tau_q$ is the cavity decay rate, $D_q=2\Delta_q/\Gamma_q$, and for reasons that will appear below, we have chosen the cavity-light detuning such that $2\delta_q/\kappa_q=D_qC_q/(1+D_q^2)$. Note that $|F_q^{i_q}|\leq1$, which is a consequence of energy conservation.

For a resonant light-atom interaction $D_q=0$, we observe that $F_q^0=1$, while $F_q^1\rightarrow-1$ for $C_q\rightarrow\infty$, i.e., the light field in the loop experiences an additional phase shift of $\pi$ if the atom is in the state $|1\rangle$ rather than $|0\rangle$. As noted in the Introduction, we obtain destructive interference in the loop if we choose $P=1$, and for $\eta_3=1$ we find $\beta^{00}=-i\alpha$, $\beta^{10}=\beta^{01}\rightarrow i\alpha$, and $\beta^{11}\rightarrow-i\alpha$, which is the optimal situation for a parity measurement. Note also that $\zeta_2^{10}=\zeta_2^{01}$ is small when $r_3^2$ is high, and this decreases the number of photons lost through spontaneous emission and other light field losses. In the nonresonant case $D_q\sim C_q\gg1$, $F_q^1\approx(F_q^0)^*$ due to the specific choice of cavity-light detuning above. The light field thus experiences one phase shift if the atom is in state $|1\rangle$ and the opposite phase shift if the atom is in state $|0\rangle$. The overall phase shift per round trip is thus zero in the odd subspace, and we choose $P=-1$ to obtain destructive interference in the loop. For $\alpha=\alpha^*$ and identical atom and cavity parameters, the conditional output field amplitudes satisfy $\beta^{10}=\beta^{01}$ and $\beta^{11}=-(\beta^{00})^*$, and a parity measurement can be achieved by detecting the $p$-quadrature of the output field. If the atom and cavity parameters are not identical and $F_1^{i}\neq F_2^{i}$, $i=0,1$, we still have $\beta^{10}=-(\beta^{01})^*$ and $\beta^{11}=-(\beta^{00})^*$ and this is sufficient to facilitate a parity measurement, but detection inefficiency now leads to decoherence within the odd subspace. The relations $\beta^{10}=-(\beta^{01})^*$ and $\beta^{11}=-(\beta^{00})^*$ are also valid during the transient as may be inferred from an argument similar to the derivation in appendix~\ref{A}.

\subsection{Validity of the approximation}\label{IID}

Having computed the steady state field amplitudes, we can now check whether assumption \eqref{condition} is consistent. The assumption is only relevant if the considered atom interacts with the light field, and for cavity 1 we thus compute $|\langle\hat{a}\rangle|^2$ conditioned on atom 1 being in state $|1\rangle$ and atom 2 being in state $|i_2\rangle$, which leads to the condition
\begin{multline}\label{newcon}
\frac{2|\alpha|^2}{\Gamma}\times
\frac{4C_1(1+D_1^2)}{(1+D_1^2+2C_1)^2+C_1^2D_1^2}\\
\times\frac{t_3^2}{|1-r_3PF_1^1F_2^{i_2}\sqrt{\eta_3}|^2}\ll1.
\end{multline}
The first factor is the number of photons in the input beam per unit time relative to the average spontaneous emission rate of an atom with an average probability of one half to be in the excited state. We anticipate that $|\alpha|^2$ is chosen such that this factor is of order unity or smaller. The second factor is small if $C_1\gg1$. In fact, this factor is of order $1/C_1$ independent of whether $D_1\sim0$, $D_1\sim1$, or $D_1\sim C_1$. In the resonant case, $D_1=0$, this appears because the possibility of spontaneous emission prevents the cavity field from building up, while in the nonresonant case, $D_1\sim C_1$, it is a consequence of the fact that the driving of the transition from $|1\rangle$ to $|e\rangle$ is inefficient when the atom-light detuning is large. The second factor is also small if $D_1\gg C_1$, but this is uninteresting, since in that limit the atom does not interact significantly with the field and $F_1^1\approx F_1^0\approx1$. Regarding the third factor, we note that $t_3$, $r_3$, $\eta_3$, $P$, $F_1^1$, and $F_2^{i_2}$ all have a norm, which is smaller than or equal to one. The factor is of order unity or smaller unless $r_3$, $\eta_3$, and $PF_1^1F_2^{i_2}$ are all close to plus one, which corresponds to the situation of constructive interference in the loop. We thus require $|1-r_3PF_1^1F_2^{i_2}\sqrt{\eta_3}|\gg C_1^{-1}$.

From a semiclassical point of view the problem in having a too large input field to cavity 1 is that the nonlinearity of the Maxwell-Bloch equations for the cavity field can no longer be neglected and this can give rise to complicated dynamics such as quantum jumps at random times between two different quasi steady states \cite{armen,armenthe}. For the same set of parameters, it is, for instance, possible to have a low amplitude of the cavity field and a low excitation of the atom or to have a large cavity field amplitude and a strong driving of the atom between $|1\rangle$ and $|e\rangle$ \cite{savage,rosenberger}. The former possibility is similar to the case of a low input field considered in the present paper, but, except for the spontaneous emission events, the latter is similar to the situation where the atom is in the state $|0\rangle$. It is then difficult to distinguish $|0\rangle$ and $|1\rangle$, and the setup is no longer suitable for a parity measurement. The condition for cavity 2 adds the requirements $C_2\gg1$ and $|1-r_3PF_1^{i_1}F_2^1\sqrt{\eta_3}|\gg C_2^{-1}$. We note that cooperativity parameters around hundred have been achieved in experiments, see for instance \cite{kimble,boca,colombe}, and the above approximation is thus realistic.

\section{Purity decay in the odd and even parity subspaces}\label{III}

As discussed in Sec.~\ref{IIA}, the probe light does not drive any transitions between the atomic states $|0\rangle$ and $|1\rangle$. On the other hand, the interaction does, in general, lead to entanglement between the atoms and the light field such that the state of the atoms and the state of the light field are correlated. A subsequent detection of the probe field may hence provide information about the state of the atoms, and this will, in general, change the coefficients $c_{i_1j_1i_2j_2}(t)$ in Eq.~\eqref{state}. In other words, the back action of the continuous measurement gradually projects the state onto an eigenstate of the measured observable, i.e., onto a state with either even or odd parity. In the ideal case, the measurement is unable to distinguish states within the odd subspace and to distinguish states within the even subspace. A main indication of the quality of the measurement is thus the rate at which a state within the odd subspace or a state within the even subspace decoheres compared to the duration of the measurement, i.e., the time required to distinguish odd parity states from even parity states.

If the atoms are initially in a state of odd parity, the state of the system at time $t$ is given by Eq.~\eqref{state} with the additional requirements $i_1\neq i_2$ and $j_1\neq j_2$. The entanglement between the atoms and the light fields reduces the purity of the atomic state during the measurement, but this purity is regained, when the input field is turned off and the cavity fields decay to the vacuum state. While the rate of nonregainable loss of purity may be slightly different during the initial and final transients, we do not expect anything dramatic to happen, since the transient dynamics does not break the symmetry, which ensures that the odd parity states are indistinguishable in the output field. For a total measurement time $t_m$, which is long compared to the transients, we may thus concentrate on the rate of loss of purity in steady state. (This is a matter of $|\alpha|^2$ being sufficiently small, since the duration of the transients does not depend on $\alpha$, while $t_m^{-1}\propto|\alpha|^2$ as we shall find below.)

In the time interval from $t$ to $t+dt$ several interactions take place. The coherent state $|\xi^{10}_1\sqrt{\tau_1}\rangle$ in cavity 1 interacts with the atom and an average number of $2C_1\kappa_1dt|\xi^{10}_1|^2\tau_1/(1+D_1^2)$ photons are lost through spontaneous emission. The loss can be modeled as a beam splitter, which transforms $|\xi^{10}_1\sqrt{\tau_1}\rangle|\textrm{vac}\rangle$ into $|\sqrt{\eta_{\textrm{SE},1}}\xi^{10}_1\sqrt{\tau_1}\rangle|
\sqrt{1-\eta_{\textrm{SE},1}}\xi^{10}_1\sqrt{\tau_1}\rangle$, where $1-\eta_{\textrm{SE},1}\equiv2C_1\kappa_1dt/(1+D_1^2)$, followed by a trace over the last mode. In steady state, $|\sqrt{\eta_{\textrm{SE},1}}\xi^{10}_1\sqrt{\tau_1}\rangle$ is transformed back into $|\xi^{10}_1\sqrt{\tau_1}\rangle$ before the next interaction with the atom. The overall transformation is thus
\begin{eqnarray}
c_{1001}\rightarrow c_{1001}\langle0|\sqrt{1-\eta_{\textrm{SE},1}}\xi_1^{10}\sqrt{\tau_1}\rangle,\\
c_{0110}\rightarrow c_{0110}\langle\sqrt{1-\eta_{\textrm{SE},1}}\xi_1^{10}\sqrt{\tau_1}|0\rangle,
\end{eqnarray}
while $c_{1100}$ and $c_{0011}$ are unchanged. Spontaneous emission in cavity 2 leads to a similar transformation, and loss through the lower mirror in Fig.~\ref{setup} gives rise to a factor $\langle\sqrt{1-\eta_3}\zeta_3^{01}\sqrt{dt}|\sqrt{1-\eta_3}\zeta_3^{10}\sqrt{dt}\rangle$ on $c_{1001}$ and the complex conjugate factor on $c_{0110}$. Finally, homodyne detection of the $p$-quadrature of the output field leaves the state unchanged if $\textrm{Im}(\beta^{10})=\textrm{Im}(\beta^{01})$. In fact, if $\beta^{10}=\beta^{01}$, it does not even lead to decoherence if the output field is simply traced out, which is why we do not consider detection inefficiency here.

Collecting the factors for the whole measurement from $0$ to $t_m$ and neglecting transients, the final purity of the atomic state after the measurement evaluates to
\begin{multline}\label{qual}
\textrm{Tr}(\rho_{\textrm{at}}(t\rightarrow\infty)^2)\approx
c_{1100}(0)^2+c_{0011}(0)^2+2|c_{1001}(0)|^2\\
\times\exp\left[-\left(\nu_{\textrm{odd},1}^{\textrm{(SE)}}
+\nu_{\textrm{odd},2}^{\textrm{(SE)}}
+\nu_{\textrm{odd}}^{\textrm{(L)}}\right)t_m\right],
\end{multline}
where $\rho_{\textrm{at}}(t)$ is the density operator of the state of the atoms obtained by tracing out the light fields,
\begin{equation}\label{nuoddse}
\nu_{\textrm{odd},q}^{\textrm{(SE)}}
=2C_q\kappa_q|\xi_q^{\delta_{1q}\delta_{2q}}|^2\tau_q/(1+D_q^2)
\end{equation}
is the rate of spontaneous emission events (i.e., the number of photons emitted spontaneously per unit time) when atom $q$ is in the state $|1\rangle$, and
\begin{equation}\label{nuoddl}
\nu_{\textrm{odd}}^{\textrm{(L)}}=(1-\eta_3)|\zeta_3^{10}-\zeta_3^{01}|^2
\end{equation}
represents the rate at which it would be possible to distinguish the two states within the odd subspace if the light lost at the lower mirror in Fig.~\ref{setup} was detected. The interpretation of Eq.~\eqref{nuoddse} follows from the fact that the right hand side is the product of the photon loss rate $\Gamma_qg_q^2/((\Gamma_q/2)^2+\Delta_q^2)$ from Eq.~\eqref{phatint} and the expectation value $|\xi_q^{\delta_{1q}\delta_{2q}}|^2\tau_q$ of the number of photons in cavity $q$, whereas the interpretation of Eq.~\eqref{nuoddl} follows from the result in appendix~\ref{B}. Additional photon losses in cavity 1 and 2 could be taken into account by including a factor of $\sqrt{\eta_q}$ in \eqref{f} and adding the terms $(1-\eta_q)|\xi_q^{10}-\xi_q^{01}|^2t_m$, $q=1,2$, to the exponent in \eqref{qual}, where $(1-\eta_q)$ is the fraction of photons lost in cavity $q$ per round trip. In appendix~\ref{B}, we show that the rate of gain of information in homodyne detection is of order the square of the distance between the conditional amplitudes of the output field projected onto the direction of the measured quadrature, and for $\alpha=\alpha^*$ and $\textrm{Im}(\beta^{10})=\textrm{Im}(\beta^{01})$, the relevant size of $t_m$ is thus
\begin{equation}\label{tm}
t_m=\textrm{max}(t_m^{00},t_m^{11}),
\end{equation}
where
\begin{equation}\label{tmii}
t_m^{ii}\equiv|\textrm{Im}(\beta^{10})-\textrm{Im}(\beta^{ii})|^{-2}.
\end{equation}

If $\textrm{Im}(\beta^{00})=\textrm{Im}(\beta^{11})$, the final purity of a state restricted to the even subspace is likewise
\begin{multline}
\textrm{Tr}(\rho_{\textrm{at}}(t\rightarrow\infty)^2)\approx
c_{1111}(0)^2+c_{0000}(0)^2+2|c_{1010}(0)|^2\\
\times\exp\left[-\left(\nu_{\textrm{even},1}^{\textrm{(SE)}}
+\nu_{\textrm{even},2}^{\textrm{(SE)}}
+\nu_{\textrm{even}}^{\textrm{(L)}}\right)t_m\right],
\end{multline}
where
\begin{equation}\label{nuevense}
\nu_{\textrm{even},q}^{\textrm{(SE)}}
=2C_q\kappa_q|\xi_q^{11}|^2\tau_q/(1+D_q^2)
\end{equation}
and
\begin{equation}\label{nuevenl} \nu_{\textrm{even}}^{\textrm{(L)}}=(1-\eta_3)|\zeta_3^{11}-\zeta_3^{00}|^2.
\end{equation}
For $\textrm{Im}(\beta^{00})\neq\textrm{Im}(\beta^{11})$, the two-qubit state is projected onto either $|00\rangle$ or $|11\rangle$ at a rate, which is of order $|\textrm{Im}(\beta^{00})-\textrm{Im}(\beta^{11})|^2\sim C^{-2}$, and we need to multiply $|c_{1111}(0)|^2$, $|c_{0000}(0)|^2$, and $|c_{1010}(0)|^2$ by the square of the accumulated weight factors arising due to the measurement.

\section{Nonresonant light-atom coupling}\label{IV}

\begin{figure}
\includegraphics[width=\columnwidth]{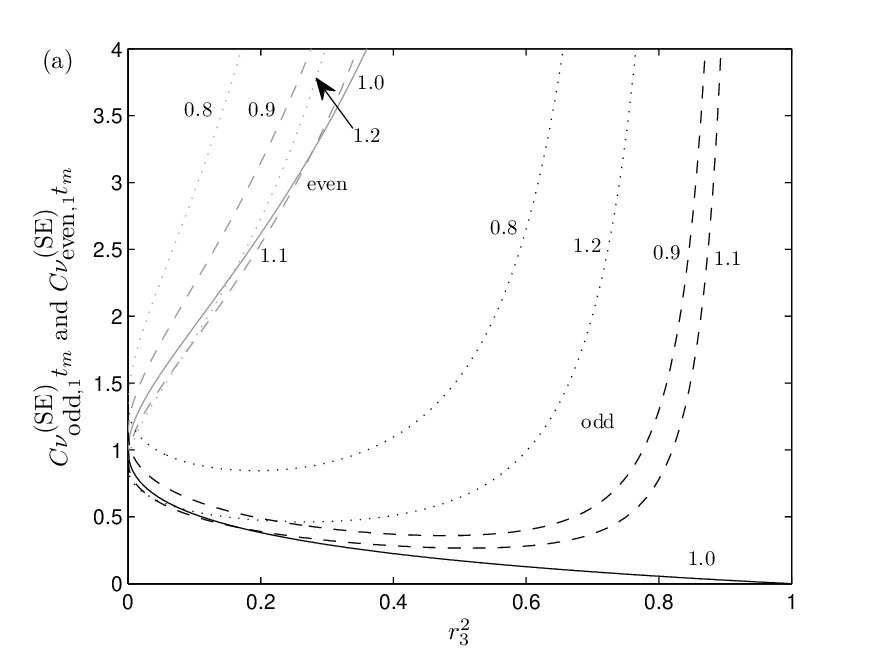}
\includegraphics[width=\columnwidth]{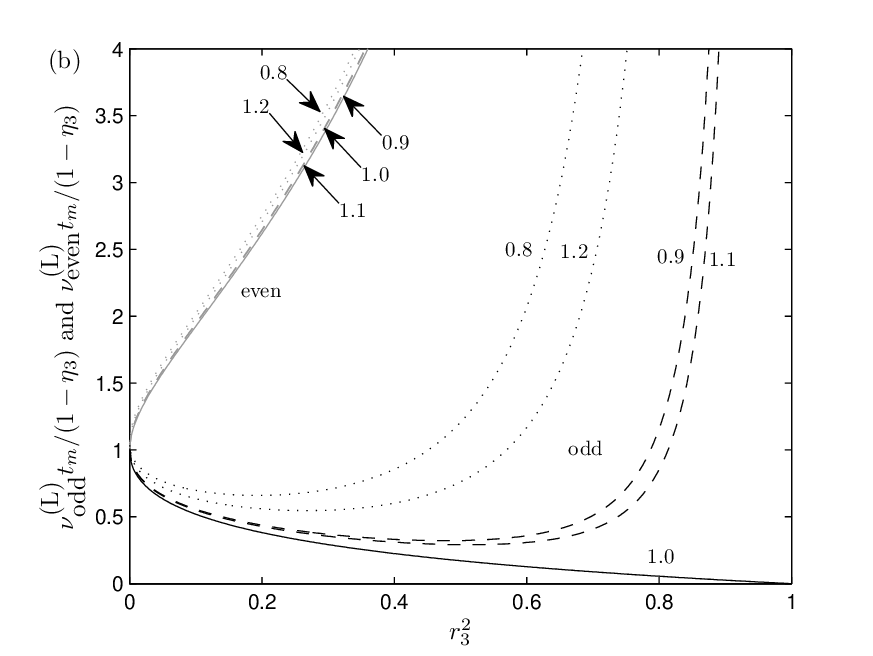}
\caption{(a) rate of spontaneous emission events relative to the rate of gain of information due to the measurement in units of $C^{-1}$ and (b) rate of decoherence within the even and odd subspaces due to loss at the lower mirror in Fig.~\ref{setup} relative to the measurement rate in units of $1-\eta_3$ for a nonresonant light-atom interaction. Both graphs are for $\eta_3=1$ and the different curves correspond to $D/C=0.8$, $0.9$, $1.0$, $1.1$, and $1.2$ as indicated. The curves in the upper left region (gray) are for the even subspace and the lower curves (black) are for the odd subspace (the curves coincide for $r_3^2=0$). \label{non}}
\end{figure}

In the following, we compare the performance of the setup in Fig.~\ref{setup} with $r_3\neq0$ and an open loop setup with $r_3=0$. We first consider a nonresonant light-atom interaction with $D\sim C$ and $C\gg1$, assuming $D_1=D_2=D$ and $C_1=C_2=C$. In this case, the primary effect of the light-atom interaction is a phase shift of the light field, and we apply the approximation $F_1^1=F_2^1\approx(D+iC)/(D-iC)\equiv F$ (see Eq.~\eqref{fqiq}). Furthermore, $F_1^0=F_2^0=F^*$. Combining $\zeta_3^{i_1i_2}=-iF_1^{i_1}\zeta_2^{i_1i_2}$, \eqref{zeta2}, \eqref{beta}, \eqref{xi}, \eqref{nuoddse}, \eqref{nuoddl}, \eqref{tm}, \eqref{tmii}, \eqref{nuevense}, and \eqref{nuevenl}, we then derive
\begin{equation}
t_m^{-1}=\frac{\eta_3(1-r_3^2)^2(1-\sqrt{\eta_3}r_3)^2(1-\textrm{Re}(F^2))^2|\alpha|^2}
{(1+\sqrt{\eta_3}r_3)^2(1+2\sqrt{\eta_3}r_3\textrm{Re}(F^2)+\eta_3r_3^2)^2},
\end{equation}
\begin{multline}
\nu^{\textrm{(SE)}}_{\textrm{odd},1}t_m=\label{nonoddse}\\
\frac{8C(1+2\sqrt{\eta_3}r_3\textrm{Re}(F^2)+\eta_3r_3^2)^2}
{\eta_3(C^2+D^2)(1-r_3^2)(1-\sqrt{\eta_3}r_3)^2(1-\textrm{Re}(F^2))^2},
\end{multline}
\begin{multline}
\nu^{\textrm{(SE)}}_{\textrm{even},1}t_m=\\
\frac{8C(1+\sqrt{\eta_3}r_3)^2
(1+2\sqrt{\eta_3}r_3\textrm{Re}(F^2)+\eta_3r_3^2)}{\eta_3(C^2+D^2)
(1-r_3^2)(1-\sqrt{\eta_3}r_3)^2(1-\textrm{Re}(F^2))^2},
\end{multline}
\begin{multline}
\nu_{\textrm{odd}}^{\textrm{(L)}}t_m=\\
\frac{4\textrm{Im}(F)^2(1-\eta_3)(1+2\sqrt{\eta_3}r_3\textrm{Re}(F^2)+\eta_3r_3^2)^2}
{\eta_3(1-\textrm{Re}(F^2))^2(1-r_3^2)(1-\sqrt{\eta_3}r_3)^2},
\end{multline}
and
\begin{equation}\label{nonevenl}
\nu_{\textrm{even}}^{\textrm{(L)}}t_m=\frac{4\textrm{Im}(F)^2(1-\eta_3)(1+\sqrt{\eta_3}r_3)^2}
{\eta_3(1-\textrm{Re}(F^2))^2(1-r_3^2)}.
\end{equation}
The latter four quantities are plotted in Fig.~\ref{non}, and we note that $\nu_{\textrm{odd},2}^{\textrm{(SE)}}=\eta_3\nu_{\textrm{odd},1}^{\textrm{(SE)}}$ and $\nu_{\textrm{even},2}^{\textrm{(SE)}}=\eta_3\nu_{\textrm{even},1}^{\textrm{(SE)}}$.

The figure shows that it is possible in the odd subspace to reduce both the rate of spontaneous emission events and the rate of decoherence due to other light field losses relative to the rate of gain of information due to the measurement by choosing a nonzero value of $r_3$, but only if $D/C$ is sufficiently close to plus or minus one, which corresponds to the situation of constructive interference in the loop for the even subspace. Note that $\nu_{\textrm{odd},q}^{\textrm{(SE)}}$ and $\nu_{\textrm{odd}}^{\textrm{(L)}}$ depend on $r_3$ in the same way, and for a given value of $D/C\in
[-\sqrt{3},-1/\sqrt{3}]\cup[1/\sqrt{3},\sqrt{3}]$ and $\eta_3=1$, the optimal choice of beam splitter reflectivity is given by
\begin{equation}
r_{3,\textrm{opt}}=\frac{\sqrt{3-3\textrm{Re}(F^2)^2}-2
-\textrm{Re}(F^2)}{1+2\textrm{Re}(F^2)}.
\end{equation}
Outside this interval, it is optimal to choose $r_3=0$. As far as light field losses in the loop are concerned, the smallest losses occur for $|D|/C=1$, and Fig.~\ref{non}(a) suggests that this is also a good choice to reduce the total number of spontaneous emission events (the optimal value of $|D|/C$ with respect to loss due to spontaneous emission varies from $\sqrt{2}$ for $\sqrt{\eta_3}r_3=0$ to $1$ for $\sqrt{\eta_3}r_3=1$). For $|D|/C\rightarrow1$, $r_{3,\textrm{opt}}\rightarrow1$, but since $\sqrt{\eta_3}PF_1^1F_2^1=+1$ for the even subspace, the results are only valid if $1-r_3\gg C^{-1}$. In fact, if $\eta_3<1$ or $C$ is not infinite, the curves go to infinity for $r_3\rightarrow1$. This happens because $t_m\rightarrow\infty$, which in turn is a consequence of the fact that it is only possible to couple light into the system for $r_3=1$ if there is perfect constructive interference within the loop and no loss, and the output field amplitude is hence $i\alpha$ for all states of the qubits if this is not the case. In conclusion, a reasonable strategy is to choose the light-atom detuning such that $|D|/C=1$, and then compute the optimal value of $r_3$ from the above expressions if the relative photon loss $1-\eta_3$ in the loop is large compared to $C^{-1}$. Otherwise one has to optimize $r_3$ under the constraint $|1+r_3F_1^1F_2^1\sqrt{\eta_3}|\gg C^{-1}$. For the even subspace, on the other hand, it is always optimal to choose $r_3=0$.

\section{Resonant light-atom coupling}\label{V}

\begin{figure}
\includegraphics[width=\columnwidth]{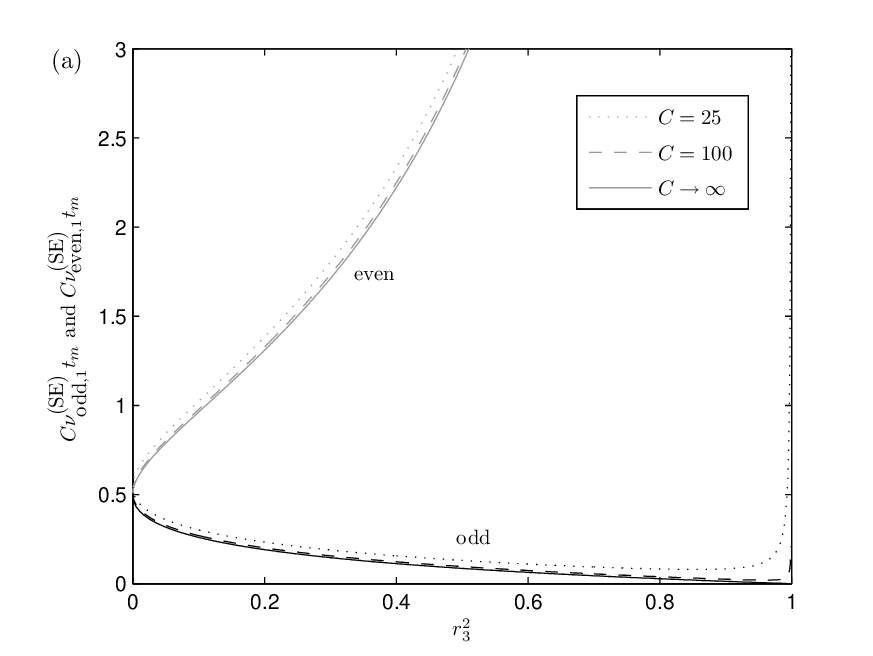}
\includegraphics[width=\columnwidth]{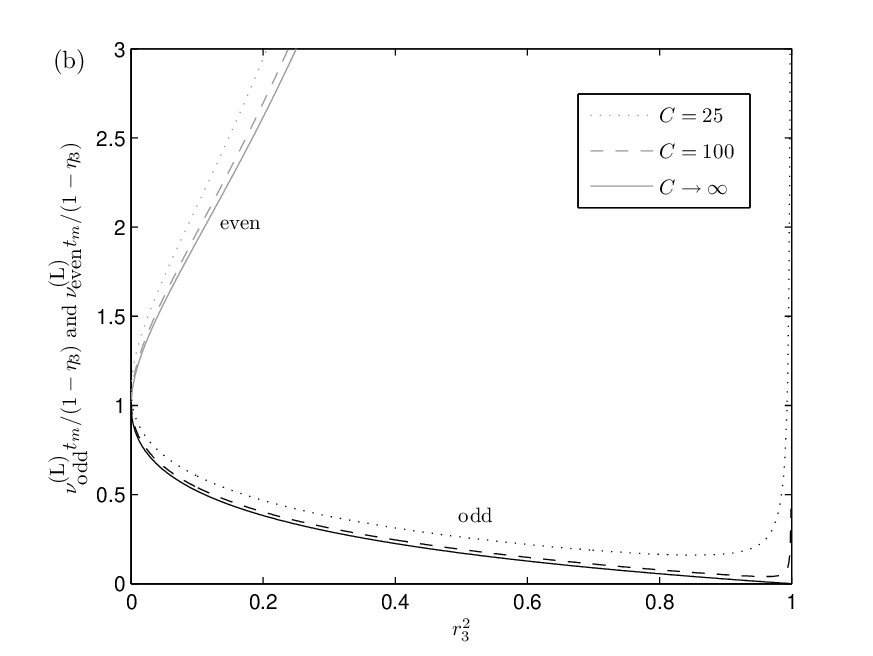}
\caption{(a) rate of spontaneous emission events relative to the measurement rate in units of $C^{-1}$ and (b) rate of decoherence within the even and odd subspaces due to loss at the lower mirror in Fig.~\ref{setup} relative to the measurement rate in units of $1-\eta_3$ for a resonant light-atom interaction and $\eta_3=1$. The solid, dashed, and dotted curves are for different values of the cooperativity parameter as shown in the legend. The curves to the upper left (gray) are for the even subspace and the lower curves (black) are for the odd subspace (the curves coincide for $r_3^2=0$). \label{res}}
\end{figure}

For a resonant light-atom coupling $D=0$ and identical atom and cavity parameters, $F_1^1=F_2^1=(1-2C)/(1+2C)\equiv G$ and $F_1^0=F_2^0=1$. In this case,
\begin{equation}
t_m^{-1}=(t_m^{11})^{-1}=\frac{\eta_3G^2(G-1)^2(1-r_3^2)^2|\alpha|^2}
{(1-\sqrt{\eta_3}r_3G)^2(1-\sqrt{\eta_3}r_3G^2)^2},
\end{equation}
\begin{equation}\label{resoddse}
\nu^{\textrm{(SE)}}_{\textrm{odd},1}t_m
=\frac{8C(1-\sqrt{\eta_3}r_3G^2)^2}{(1+2C)^2\eta_3G^2(G-1)^2(1-r_3^2)},
\end{equation}
\begin{equation}
\nu^{\textrm{(SE)}}_{\textrm{even},1}t_m
=\frac{8C(1-\sqrt{\eta_3}r_3G)^2}{(1+2C)^2\eta_3G^2(G-1)^2(1-r_3^2)},
\end{equation}
\begin{equation}
\nu_{\textrm{odd}}^{\textrm{(L)}}t_m=
\frac{(1-\eta_3)(1-\sqrt{\eta_3}r_3G^2)^2}{\eta_3G^2(1-r_3^2)},
\end{equation}
and
\begin{equation}\label{resevenl}
\nu_{\textrm{even}}^{\textrm{(L)}}t_m=\frac{(1-\eta_3)(1-\eta_3r_3^2G^2)^2}
{\eta_3G^2(1-r_3^2)(1-\sqrt{\eta_3}r_3)^2},
\end{equation}
while $\nu_{\textrm{odd},2}^{\textrm{(SE)}}=\eta_3\nu_{\textrm{odd},1}^{\textrm{(SE)}}$ and $\nu_{\textrm{even},2}^{\textrm{(SE)}}=\eta_3G^2\nu_{\textrm{even},1}^{\textrm{(SE)}}$. These results are illustrated for $\eta_3=1$ in Fig.~\ref{res}, which shows that the total decoherence in the odd subspace also in this case can be reduced by choosing a nonzero $r_3$. For a given value of $C$ and $\eta_3$, the minimum appears at $r_{3,\textrm{opt}}=\sqrt{\eta_3}G^2$, but again one should keep the constraint $|1-r_3F_1^1F_2^1\sqrt{\eta_3}|\gg C^{-1}$ in mind.

Assuming $D=C$ and $C\rightarrow\infty$ for the nonresonant case and $C\rightarrow\infty$ for the resonant case, we observe that the expressions for $t_m$ are the same, and the two possibilities are thus equally efficient in distinguishing the even and the odd subspaces. We also note that $\zeta_1^{i_1i_2}$ is the same for the resonant and the nonresonant coupling, leading to the same values for the rate of decoherence due to light field losses in the loop. The rates of spontaneous emission are also the same except for a factor of 2, which arises from the details of the light-atom interaction: For cavities containing an atom in the state $|1\rangle$, the number of photons in the cavity is a factor of $1/(2C^2)$ smaller in the resonant case compared to the nonresonant case, but the fraction of photons lost per round trip is a factor of $C^2$ larger. Finally, the significance of additional photon losses in cavity 1 and 2 is the same for resonant and nonresonant coupling because $|\xi_q^{10}-\xi_q^{01}|$ is the same and $|\xi_q^{11}-\xi_q^{00}|$ is the same. The performance is thus roughly the same for a resonant and a nonresonant light-atom interaction.

It is interesting to note that the total number of spontaneous emission events is of order $C^{-1}$, while the contribution to the exponent in \eqref{qual} from loss at the lower mirror in Fig.~\ref{setup} is of order $(1-\eta_3)$, such that the relative importance of the two effects is determined by $C(1-\eta_3)$. The importance of additional photon losses in cavity 1 or 2 relative to spontaneous emission is likewise given by $C(1-\eta_q)$. High performance is thus obtained for high cooperativity parameters and low propagation and reflection losses for the light field. On the other hand, Eqs.~\eqref{nonoddse}-\eqref{nonevenl} and Eqs.~\eqref{resoddse}-\eqref{resevenl} show that the performance is independent of the intensity of the input field as long as Eq.~\eqref{newcon} is satisfied. This is the case, because both the rate of gain of information due to the measurement and the photon loss rates scale linearly with $|\alpha|^2$. Equation~\eqref{newcon} defines an upper limit for $|\alpha|^2$, but $|\alpha|^2$ should not be chosen too small either in order to limit the time required to carry out the parity measurement.

\section{Conclusion}\label{VI}

In conclusion, we have analyzed the performance of a parity measurement on two qubits, each encoded in two ground state levels of a single atom in a cavity. The measurement relies on the phase shift imposed conditionally on a continuous beam of light, which interacts sequentially with the two qubits. We have found that the rate of decoherence within the odd parity subspace due to spontaneous emission and light field losses can be decreased by allowing the probe field to interact several times with the two qubits before it is detected, but the decrease happens at the expense of an increased decoherence rate within the even parity subspace. Finally, we have shown that the performance of the parity measurement is about the same for a resonant and for a nonresonant light-atom interaction if the light-atom detuning is chosen as $|\Delta|=g^2/\kappa$, where $g$ is the light-atom coupling strength and $\kappa$ is the decay rate of light in the cavity.

The proposed measurement can, for instance, be used to prepare two qubits in an entangled odd parity state with higher purity. Improvements can also be obtained in quantum error correction if the qubit state is encoded in the odd subspace and errors are corrected continuously such that the even subspace is practically avoided. In case of the three-qubit bit-flip correction code, for instance, one would use the three-qubit states $|010\rangle$ and $|101\rangle$ as the code space and then perform the proposed parity measurement on qubits 1 and 2 and on qubits 2 and 3. This could be done either by switching between the two measurements or by using two different polarizations of the light field, one of which interacts with the qubit state $|1\rangle$ and the other with the qubit state $|0\rangle$. Polarizing beam splitters could then be used to guide the polarization components into the right cavities.

We have presented the proposed structure as a closed loop setup, but we note the similarities with both a coherent feedback loop and a cavity array. The idea of having cavities within cavities and utilizing constructive and destructive interference may be useful in other settings as well.

\appendix

\section{Symmetry properties of the output field}\label{A}

We show that the output field is the same for the two odd parity states even if the input field is time dependent, provided the cavity and atom parameters are the same for the two cavities. At first, we only require $\tau_1=\tau_2\equiv\tau$. We denote the light traveling time from cavity 1 to cavity 2 by $T$ and choose the time $t=0$ such that $\zeta_2^{i_1i_2}(t<0)=0$, while $\zeta_2^{i_1i_2}(t)$ may have an arbitrary time dependence for $t\geq 0$. Inspection of Fig.~\ref{setup} then gives
\begin{multline}\label{symout}
\zeta_5^{i_1i_2}(T+(n+x)\tau)=
\sum_{q=0}^{n-1}\Bigg( it_1^2r_2f_1^{i_1}\sqrt{\eta_3}(r_1f_1^{i_1})^{n-q-1}\\
+ir_1t_2^2f_2^{i_2}\sqrt{\eta_3}(r_2f_2^{i_2})^{n-q-1}
-it_1^2t_2^2f_1^{i_1}f_2^{i_2}\sqrt{\eta_3}\\
\times\sum_{k=0}^{n-q-2}(r_1f_1^{i_1})^k(r_2f_2^{i_2})^{n-q-k-2}\Bigg)\zeta_2^{i_1i_2}((q+x)\tau)\\
-ir_1r_2\sqrt{\eta_3}\zeta_2^{i_1i_2}((n+x)\tau),
\end{multline}
where $n\in\mathbb{N}_0$ and $x\in[0,1[$. The first term in the brackets is the contribution from the field, which enters into cavity 1 at time $(q+x)\tau$ and travels $n-q$ round trips before it leaves cavity 1 and travels to the input mirror of cavity 2 where it is reflected. The second term is the contribution from the field, which travels zero round trips in cavity 1 and $n-q$ round trips in cavity 2, and the third term is the contribution from the field, which travels $k+1$ round trips in cavity 1 and $n-q-k-1$ round trips in cavity 2. The final term is the contribution from the field, which travels zero round trips in both cavity 1 and 2.

For $r_1=r_2$ and $t_1=t_2$, we note that $\zeta_5^{i_1i_2}$ is invariant under exchange of $i_1$ and $i_2$ if $\zeta_2^{i_1i_2}$ is invariant at all times and $f_1^0=f_2^0$ and $f_1^1=f_2^1$. The former requirement is fulfilled if $\zeta_5^{i_1i_2}$ is invariant, since $\zeta_2^{i_1i_2}$ is a linear combination of $\alpha$ and $\zeta_5^{i_1i_2}$, and the latter requirement is fulfilled if the cavity and atom parameters are the same for the two cavities. The invariance of $\beta^{i_1i_2}$ then follows from the fact that $\beta^{i_1i_2}$ is a linear combination of $\alpha$ and $\zeta_5^{i_1i_2}$.

\section{Rate of gain of information due to homodyne detection}\label{B}

We estimate the ability of homodyne detection to distinguish continuous coherent states. Consider the light field mode, which is detected in the time interval from $t$ and $t+dt$. We define the integrated photo current as $dy\equiv k/|\alpha|$, where $k$ is the difference in the observed number of photons at the two detectors within the time interval and $|\alpha|^2$ is the average number of photons per unit time in the local oscillator beam. Following the derivation in \cite{nm6}, and assuming that the measured light field mode is in the coherent state $|\beta\sqrt{dt}\rangle$, we find
\begin{equation}
dy=dW-ie^{-i\theta}\beta dt+ie^{i\theta}\beta^*dt,
\end{equation}
where $dW$ is a Gaussian stochastic variable with mean zero and variance $dt$ and $\theta$ is the phase of the local oscillator. The mean value of the photo current is thus $\langle\langle dy\rangle\rangle=-ie^{-i\theta}\beta dt+ie^{i\theta}\beta^*dt$, and the standard deviation is $\sqrt{dt}$.

The measurement time $t_m$ required to distinguish two coherent states with amplitudes $\beta_1$ and $\beta_2$ may now be defined as the time for which the absolute value of the difference in the mean value of the photo current integrated from $0$ to $t_m$
\begin{multline}
\left|\int_0^{t_m}\langle\langle dy_1\rangle\rangle-\int_0^{t_m}\langle\langle dy_2\rangle\rangle\right|=|-ie^{-i\theta}\beta_1t_m\\
+ie^{i\theta}\beta_1^*t_m+ie^{-i\theta}\beta_2 t_m-ie^{i\theta}\beta_2^*t_m|
\end{multline}
equals twice the standard deviation $2\sqrt{t_m}$, i.e.,
\begin{equation}\label{nym} t_m=4|-ie^{-i\theta}\beta_1+ie^{i\theta}\beta_1^*+ie^{-i\theta}\beta_2 -ie^{i\theta}\beta_2^*|^{-2}.
\end{equation}
The rate of gain of information is thus the square of the distance in phase space between the two coherent state amplitudes projected onto the direction of the measured quadrature.

\begin{acknowledgments}
The author acknowledges discussions with Joseph Kerckhoff, Hendra Nurdin, and Hideo Mabuchi and financial support from the Danish Minister of Science, Technology, and Innovation.
\end{acknowledgments}


\begin{thebibliography}{99}
\bibitem{knill} E. Knill, R. Laflamme, and G. Milburn, Nature (London) {\bf409}, 46 (2001).
\bibitem{ourjoumtsev} A. Ourjoumtsev, R. Tualle-Brouri, J. Laurat, and P. Grangier, Science {\bf312}, 83 (2006).
\bibitem{neergaard} J. S. Neergaard-Nielsen, B. M. Nielsen, C. Hettich, K. M{\o}lmer, and E. S. Polzik, Phys. Rev. Lett. {\bf97}, 083604 (2006).
\bibitem{wakui} K. Wakui, H. Takahashi, A. Furusawa, and M. Sasaki, Opt. Express {\bf15}, 3568 (2007).
\bibitem{kuzmich} A. Kuzmich, L. Mandel, and N. P. Bigelow, Phys. Rev. Lett. {\bf85}, 1594 (2000).
\bibitem{massar} S. Massar and E. S. Polzik, Phys. Rev. Lett. {\bf91}, 060401 (2003).
\bibitem{nm8} A. E. B. Nielsen, U. V. Poulsen, A. Negretti, and K. M{\o}lmer, Phys. Rev. A {\bf79}, 023841 (2009).
\bibitem{gottesman} D. Gottesman, e-print arXiv:0904.2557.
\bibitem{williams} N. S. Williams and A. N. Jordan, Phys. Rev. A {\bf78}, 062322 (2008).
\bibitem{ionicioiu} R. Ionicioiu, Phys. Rev. A {\bf75}, 032339 (2007).
\bibitem{yuan} Q. Yuan and J. Li, Sci China Ser G {\bf52}, 1203 (2009).
\bibitem{kerckhoff} J. Kerckhoff, L. Bouten, A. Silberfarb, and H. Mabuchi, Phys. Rev. A {\bf79}, 024305 (2009).
\bibitem{hood} C. J. Hood, M. S. Chapman, T. W. Lynn, and H. J. Kimble, Phys. Rev. Lett. {\bf80}, 4157 (1998).
\bibitem{duan} L. M. Duan and H. J. Kimble, Phys. Rev. Lett. {\bf92},
    127902 (2004).
\bibitem{ahn} C. Ahn, A. C. Doherty, and A. J. Landahl, Phys. Rev. A {\bf65}, 042301 (2002).
\bibitem{sarovar} M. Sarovar, C. Ahn, K. Jacobs, and G. J. Milburn, Phys. Rev. A {\bf69}, 052324 (2004).
\bibitem{nurdin} J. Kerckhoff, H. I. Nurdin, D. S. Pavlichin, and H. Mabuchi, e-print arXiv:0907.0236.
\bibitem{bouten1} L. Bouten and A. Silberfarb, Commun. Math. Phys. {\bf283}, 491 (2008).
\bibitem{bouten2} L. Bouten, R. van Handel, and A. Silberfarb, Journal of Functional Analysis {\bf254}, 3123 (2008).
\bibitem{yanagisawa} M. Yanagisawa and H. Kimura, IEEE Trans. automatic control {\bf48}, 2107 (2003).
\bibitem{bouten3} L. Bouten, R. van Handel, and M. R. James, SIAM J.
    Control Optim. {\bf46}, 2199 (2007).
\bibitem{blow} K. J. Blow, R. Loudon, S. J. D. Phoenix, and T. J. Shepherd, Phys. Rev. A {\bf42}, 4102 (1990).
\bibitem{giedke} G. Giedke and J. I. Cirac, Phys. Rev. A {\bf66}, 032316 (2002).
\bibitem{armen} M. A. Armen and H. Mabuchi, Phys. Rev. A {\bf73}, 063801 (2006).
\bibitem{armenthe} M. Armen, Bifurcations in single atom cavity QED, PhD Thesis, California Institute of Technology (2009).
\bibitem{savage} C. M. Savage and H. J. Carmichael, IEEE J. Quant. Elec. {\bf24}, 1495 (1988).
\bibitem{rosenberger} A. T. Rosenberger, L. A. Orozco, H. J. Kimble, and P. D. Drummond, Phys. Rev. A {43}, 6284 (1991).
\bibitem{kimble} H. J. Kimble, Phys. Scr. {\bf T76}, 127 (1998).
\bibitem{boca} A. Boca, R. Miller, K. M. Birnbaum, A. D. Boozer, J. McKeever, and H. J. Kimble, Phys. Rev. Lett. {\bf93}, 233603 (2004).
\bibitem{colombe} Y. Colombe, T. Steinmetz, G. Dubois, F. Linke, D. Hunger, and J. Reichel, Nature (London) {\bf450}, 272 (2007).
\bibitem{nm6} A. E. B. Nielsen and K. M{\o}lmer, Phys. Rev. A {\bf77}, 052111 (2008).
\end{thebibliography}
\end{document}